\documentclass{article}

\usepackage[preprint]{main}
\usepackage{xspace}

\newcommand{\AttackName}{$\text{M}^2\text{A}$\xspace}
\newcommand{\SourceCode}{https://github.com/Momoyeyu/M2A}

\newcommand{\etal}{{\textit{et al.}}}

\usepackage{graphicx}
\usepackage{subcaption}     
\usepackage{float}
\usepackage{algorithm}
\usepackage{algpseudocode}

\usepackage[utf8]{inputenc} 
\usepackage[T1]{fontenc}    
\usepackage{hyperref}       
\usepackage{url}            
\usepackage{booktabs}       
\usepackage{amsfonts}       
\usepackage{nicefrac}       
\usepackage{microtype}      
\usepackage{xcolor}         
\usepackage{multirow}
\usepackage{amsmath}
\usepackage{enumitem}
\usepackage{tcolorbox}
\tcbset{
  colback=gray!10,        
  colframe=black!60,     
  coltitle=white,          
  colbacktitle=black!60, 
  fonttitle=\bfseries,   
  boxrule=0.5pt,
  arc=3pt,
  left=5pt,
  right=5pt,
  top=5pt,
  bottom=5pt
}

\title{Mirage Fools the Ear, Mute Hides the Truth: Precise Targeted Adversarial Attacks on Polyphonic Sound Event Detection Systems}

\author{
Junjie Su$^1$, \,
Weifei Jin$^1$, \,
Yuxin Cao$^2$, \,
Derui Wang$^3$, \,
Kai Ye$^4$, \,
Jie Hao$^1$\thanks{Corresponding author} \\
$^1$Beijing University of Posts and Telecommunications \\
$^2$National University of Singapore \,
$^3$CSIRO’s Data61 \, 
$^4$The University of Hong Kong \\
\texttt{\{sujunjie, haojie\}@bupt.edu.cn}
}

\begin{document}

\maketitle

\begin{abstract}
Sound Event Detection (SED) systems are increasingly deployed in safety-critical applications such as industrial monitoring and audio surveillance. However, their robustness against adversarial attacks has not been well explored. Existing audio adversarial attacks targeting SED systems, which incorporate both detection and localization capabilities, often lack effectiveness due to SED's strong contextual dependencies or lack precision by focusing solely on misclassifying the target region as the target event, inadvertently affecting non-target regions. To address these challenges, we propose the Mirage and Mute Attack (\AttackName) framework, which is designed for targeted adversarial attacks on polyphonic SED systems. In our optimization process, we impose specific constraints on the non-target output, which we refer to as \textit{preservation loss}, ensuring that our attack does not alter the model outputs for non-target region, thus achieving precise attacks. Furthermore, we introduce a novel evaluation metric Editing Precison (EP) that balances effectiveness and precision, enabling our method to simultaneously enhance both. Comprehensive experiments show that \AttackName achieves 94.56\% and 99.11\% EP on two state-of-the-art SED models, demonstrating that the framework is sufficiently effective while significantly enhancing attack precision.
\end{abstract}

\section{Introduction}
\label{sec:intro}

Nowadays, automated surveillance systems have significantly improved in detecting real-time threats \citep{dong2020automatic, yatbaz2024run}. While video-based surveillance is widely used, audio-based systems provide a cost-effective, privacy-conscious alternative. They eliminate blind spots, process signals efficiently, and avoid capturing sensitive visual data \citep{clavel2005events, crocco2016audio}. These advantages make audio surveillance particularly valuable in industrial monitoring \citep{grollmisch2019sounding, shreyas2020trends, hikvision}, security \citep{valenzise2007scream, neri2022sound}, healthcare \citep{naimul2023medical}, and smart cities \citep{staahlbrost2014audio}.

Modern Sound Event Detection (SED) systems analyze complex auditory scenes to identify multiple overlapping sounds with temporal precision. This polyphonic capability is essential for industrial automation and smart surveillance, where accurate event localization is critical \citep{crocco2016audio}. However, despite their widespread adoption, the security of SED systems remains underexplored, posing significant risks. For example, an attacker could exploit these vulnerabilities to bypass an SED-based security system during a break-in, preventing detection of critical sounds like forced door openings or glass breaking. This failure could delay security responses, allowing intruders unauthorized access and causing severe consequences, including financial loss or physical harm. Existing adversarial attack methods, primarily designed for speech-related tasks, fail to account for SED’s strong temporal dependencies, often causing imprecise attacks that introduce unintended modifications to non-target events. To address this gap, we propose the first targeted adversarial attack framework specifically designed for SED, enabling precise event manipulations while maintaining stealthiness.

\begin{figure}[t]
\centering
\includegraphics[width=0.70\linewidth]{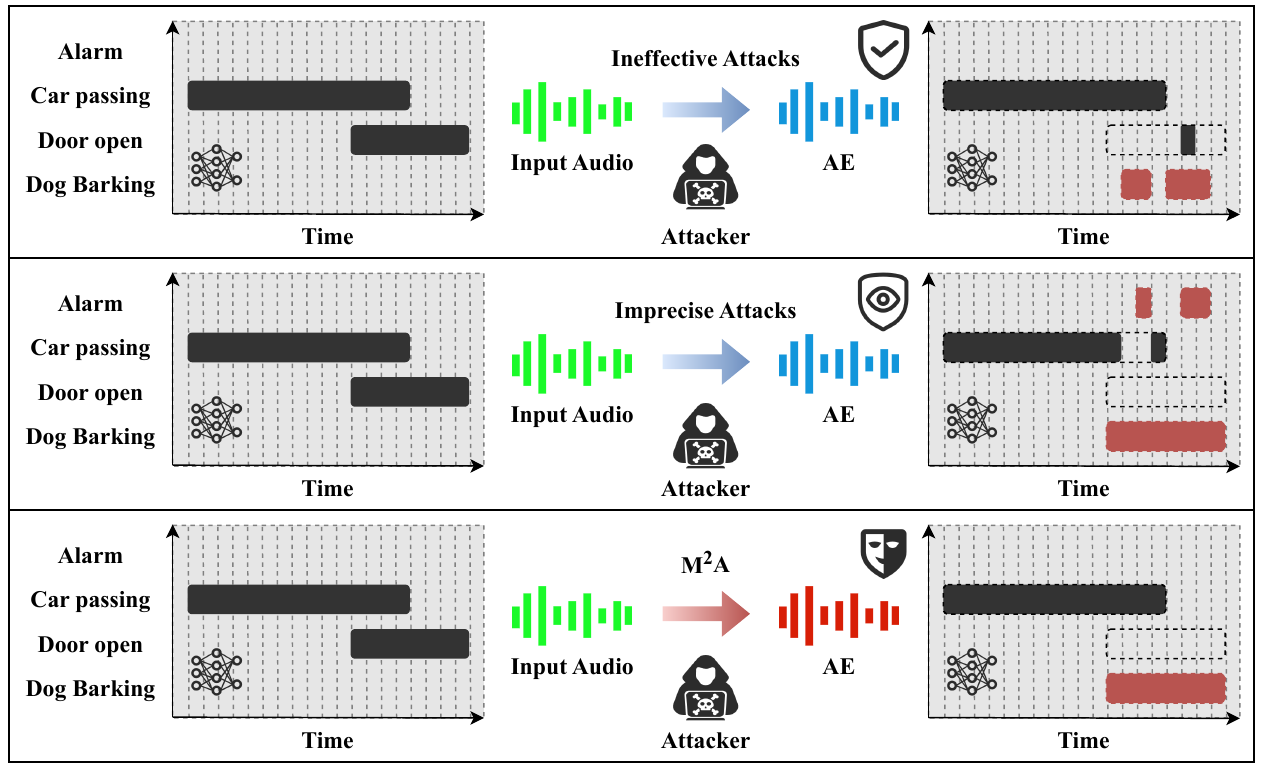}
\caption{All attacks aim to delete `door open' and insert `dog barking'. Ineffective attacks fail to fully achieve their goal, while imprecise ones disrupt the non-target region. Our method ensures effectiveness while better preserving the non-target region.
}
\label{fig:attack_overview}
\end{figure}

Adversarial attacks \citep{szegedy2013intriguing, goodfellow2014explaining} on audio systems are well documented in speech-related tasks \citep{carlini2018audio, yakura2018robust, miao2022faag, qu2022synthesising, jin2024towards}. However, these approaches are inadequate for SED. As illustrated in Figure~\ref{fig:attack_overview}, when applied to SED, these methods either fail to modify the target region effectively (lack of effectiveness) due to the strong context dependencies inherent in SED \citep{heittola2013context}, or inadvertently alter the non-target region (lack of precision) due to excessive perturbations or an inadequate consideration for SED tasks, resulting in implausible outputs that are easily detected \citep{yurdakul2023acoustic}. Such weaknesses significantly reduce the stealthiness of adversarial examples.

To improve both attack effectiveness and precision, we propose \textbf{M}irage and \textbf{M}ute \textbf{A}ttack (\AttackName), an adversarial attack framework that supports the insertion (Mirage) or deletion (Mute) of targeted events. The name encapsulates the core requirements of our approach: (1) \textbf{Effectiveness}: The perturbations successfully modify model output in the target region to align with the attacker's intent. (2) \textbf{Precision}: The unintended modifications are minimized to prevent the attack from being detected. We perturb only targeted segments to enhance stealthiness while considering global audio context for attack effectiveness. To minimize influence on the non-target region, we introduce a \textit{preservation loss} constraint, ensuring output consistency and reducing unintended impact. Furthermore, our method supports arbitrary manipulations, allowing multiple events in the target region to be modified simultaneously, which means that our method has high practicality.

In addition, since existing evaluation metrics mostly focus on attack effectiveness and audio quality, they lack an assessment of attack precision. To provide a more comprehensive evaluation of our method, we propose a new evaluation metric specifically designed for attacking SED tasks: \textbf{Editing Precision (EP)}, which jointly quantifies the Attack Success Rate (ASR) and the Unintended Editing Rate (UER). Using this metric, we conduct extensive experimental evaluations on two widely used datasets, TUT-SED~\citep{mesaros2017tut} and DESED~\citep{turpault2019sound}, with state-of-the-art models, CRNN~\citep{adavanne2017sound} and ATST-SED~\citep{shao2024fine}. Our method achieves an EP of \textbf{94.56\% on CRNN and 99.11\% on ATST-SED}, demonstrating high attack effectiveness and precision.
The source code and demos are anonymously available at: \url{\SourceCode}.

In summary, our contributions are as follows:

\begin{itemize}[leftmargin=*]
    \item \textbf{First Adversarial Attack Framework for Polyphonic SED}: To the best of our knowledge, we are the first to exploit adversarial attacks on polyphonic SED. We propose \AttackName framework, which enables precise event manipulations.
    \item \textbf{Precise Attack with Preservation Loss}: We propose a novel preservation loss constraint that reduces unintended manipulations in non-target regions, which improves attack precision and in turn enhances the stealthiness of the attack behavior.
    \item 
    \textbf{Extensive Evaluation}: Experiments on two widely-used datasets and two state-of-the-art SED models reveal that our method can achieve outstanding attack effectiveness and high precision.
\end{itemize}

\section{Preliminaries and Related Work}

\subsection{Sound Event Detection}

Unlike Sound Event Classification (SEC), which focuses solely on labeling events, Sound Event Detection (SED) involves temporal localization, determining their precise start and end times. This makes SED a more complex and critical task for applications such as environmental monitoring, security surveillance, and industrial monitoring.

SED can be categorized into monophonic SED and polyphonic SED \citep{mesaros2016metrics, chan2020comprehensive}. Monophonic SED involves detecting single, non-overlapping sound events, while polyphonic SED handles multiple overlapping events occurring simultaneously within an audio clip. We focus on polyphonic SED since it is more practical for audio surveillance scenarios.

Given an SED model $f:\mathbb{R}^{N}\rightarrow \{0,1\}^{N\times C} $ and an input audio $x \in \mathbb{R}^{N}$, $f$ first processes $x$ into a mel-spectrum and generates frame-level event predictions:
\begin{equation}
    \hat{y} = f(x), \hat{y} \in \{0,1\}^{N \times C},
\end{equation}
where $C$ is the number of sound event categories and $N$ is the temporal dimension.

Figure~\ref{fig:sed-output} presents polyphonic SED output as a frame-wise event activity matrix, where the horizontal axis represents time frames and the vertical axis corresponds to event classes. Each predicted value $\hat{y}_{t,e}$ indicates the confidence of event $e$ occurring at time $t$, which is thresholded to obtain binary event activations.

The model is trained on a dataset of size $M$ using a binary cross-entropy (BCE) loss:
\begin{equation}
\mathcal{L} = \frac{1}{M} \sum_{i=1}^{M} \sum_{t=1}^{N} \sum_{e=1}^{C} \ell_{\text{BCE}}(\hat{y}_{t,e}^{(i)}, y_{t,e}^{(i)}),
\end{equation}
where $y_{t,e}^{(i)}$ represents the ground truth event occurrence at time $t$ for event class $e$ in the $i$-th training sample of the dataset.

The evolution of SED has progressed from handcrafted feature-based models like frame-based CRNNs \citep{adavanne2017sound} to more scalable solutions \citep{mesaros2021sound}. Mobile-friendly CRNNs optimized via model distillation enable real-time applications \citep{fu2019mobile}, while Transformer-based models eliminate manual feature extraction and segmentation \citep{ye2021sound}. Recent innovations, such as DiffSED, enhance robustness in noisy environments using denoising diffusion \citep{bhosale2024diffsed}. Among state-of-the-art models, ATST-SED demonstrates superior performance in polyphonic SED tasks \citep{shao2024fine}, reflecting the field's rapid advancement. 

\subsection{Audio Adversarial Attacks}

The majority of audio adversarial attacks have been developed for speech-related tasks, such as automatic speech recognition and speaker verification. Carlini \etal~\cite{carlini2018audio} proposed optimization-based methods to attack speech-to-text models in a white-box setting, demonstrating effectiveness against popular speech-to-text systems. Other approaches, such as Y\&S~\cite{yakura2018robust}, focused on enhancing the robustness of adversarial examples (AEs), particularly under physical-world transformations. FAAG~\cite{miao2022faag} aimed to improve computational efficiency, reducing the cost of generating AEs. Additionally, some methods leverage auxiliary models to generate AEs; for instance, SSA~\cite{qu2022synthesising} employs a text-to-speech model to synthesize AEs when original audio is unavailable, while STA~\cite{jin2024towards} incorporates a style transfer model. More recent work~\cite{chen2023qfa2sr, ge2023advddos, fang2024zero, jin2025boosting, jin2025whispering} has explored the universality and transferability of AEs across different models and settings.

Subramanian \etal \cite{subramanian2019adversarial} applied several existing adversarial attack methods~\cite{goodfellow2014explaining, moosavi2016deepfool, carlini2018audio} to Sound Event Classification (SEC) tasks. They further explored the transferability of AEs against SEC models to uncover more fundamental vulnerabilities of SEC models to adversarial manipulations \citep{subramanian2020study}. These studies marked an initial exploration of adversarial attacks in the context of environmental sound safety. However, their methods were tested under relatively simple conditions, which fail to fully capture the complexities of audio surveillance based on polyphonic SED systems, where events need to be precisely targeted without disturbing unrelated sounds. There remains a significant gap in research regarding the robustness of SED systems, and we aim to explore this gap by investigating the vulnerability of polyphonic SED systems to adversarial attacks.

\section{Methodology}

This section formalizes the adversarial attack problem for SED, noting its differences from other tasks. We present the proposed preservation loss constraint, detailing its rationale and mechanism, followed by a thorough description of the attack algorithm and its optimization process.

\begin{minipage}{0.48\textwidth}
  \begin{figure}[H]
    \centering
    \includegraphics[width=\linewidth]{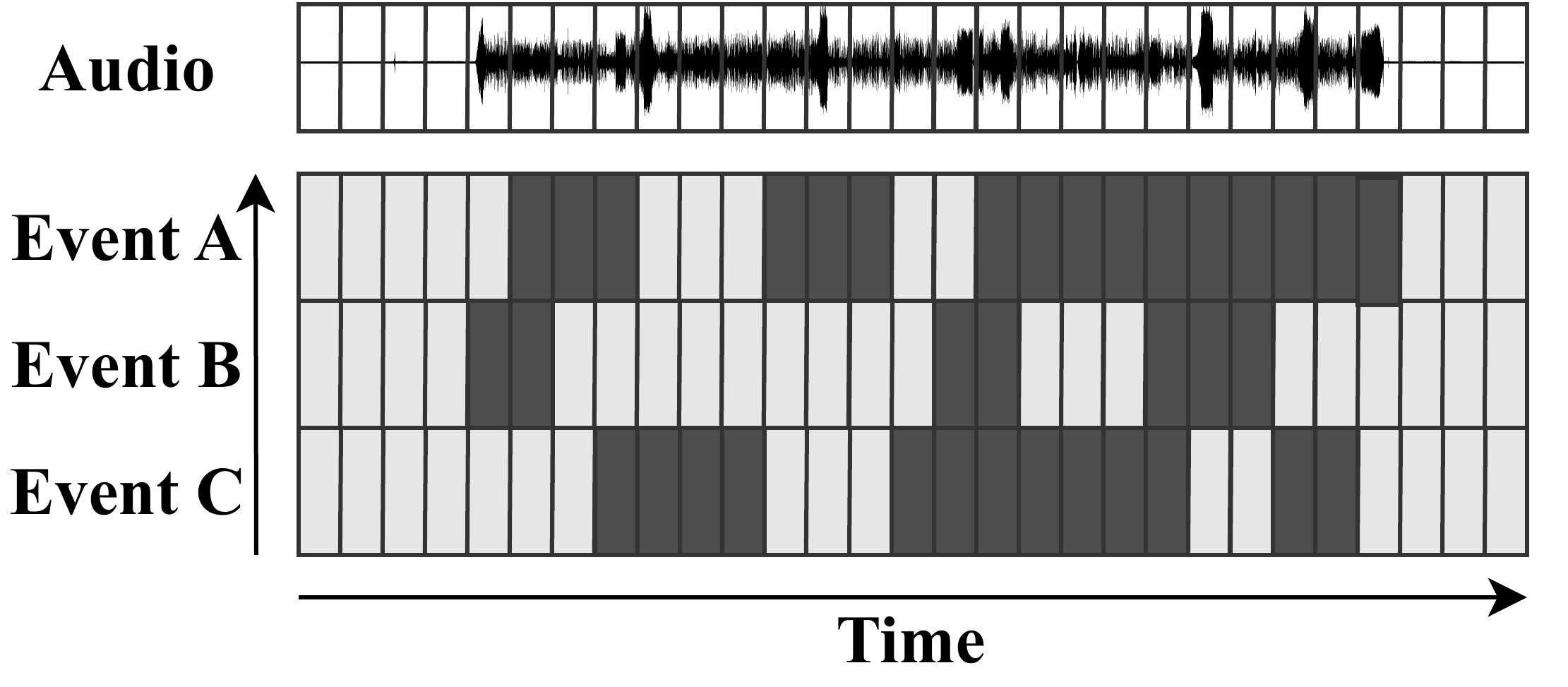}
    \caption{Polyphonic SED output: The audio is divided into multiple frames and each frame predicts the presence of different sound events.}
    \label{fig:sed-output}
  \end{figure}
\end{minipage}%
\hfill
\begin{minipage}{0.48\textwidth}
  \begin{figure}[H]
      \hfill
      \includegraphics[width=\linewidth]{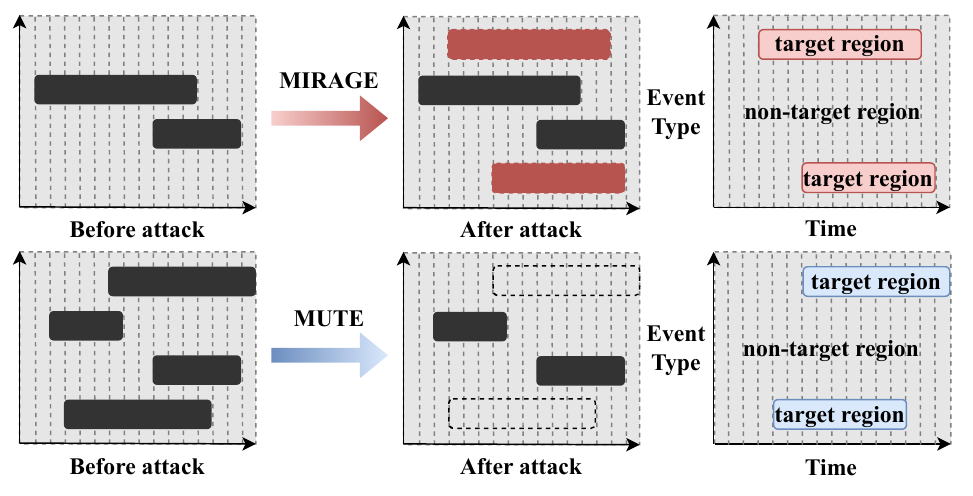}
      \caption{Visualization of the target region and the non-target region under \AttackName framework.}
      \label{fig:target-region}
    \end{figure}
\end{minipage}

\subsection{Problem Formulation}\label{sec:formulation}
Figure~\ref{fig:target-region} illustrates the definition of target and non-target regions. Let the target SED model be $f:\mathbb{R}^{N}\rightarrow \{0,1\}^{N\times C} $, which takes an input audio $x \in \mathbb{R}^{N}$ and outputs $\hat{y} = f(x),  \hat{y} \in \{0,1\}^{N \times C}$, where \( N \) is the temporal dimension and \( C \) is the number of event types that the SED model can recognize. Given the output index set $O=\{(i,j)|\,i\in[1,N], \, j\in [1,C]\}$, for each time-event pair $(i,j)\in O$, \( f(x)_{i,j} \) is 0 or 1, indicating whether an event occurs at a given time step.

In adversarial attacks, the adversary's goal is defined as:
\begin{equation}
\begin{aligned}
f(x+\delta)=y^* \quad \text{s.t.} \, \, \|\delta\|_{\infty}\leq \tau ,
\end{aligned}
\end{equation}
where \( \delta \) is the adversarial perturbation constrained under a budget $\tau$, $y^*$ is the target label matrix. Given each target value $t \in \{0,1\}$ (0 if deletion, 1 if insertion), $y^*$ can be further defined as:
\begin{equation}
\begin{aligned}
&\forall (i,j) \in T \subset O,\,\, y^*_{i,j} = t \\
&\forall (i,j) \in O \setminus  T, \,\, y^*_{i,j} = \hat{y}_{i,j},
\end{aligned}
\end{equation}
where $T$ is the index set of the target region. Given the target temporal interval $[l,r]$ and the target event $e$, the target region $U=\{(l,e),(l+1,e),\dots,(r-1,e),(r,e)\}$. For single-target manipulation, $T=U$, while for multi-target manipulations, given multiple target regions $\{U_i\}$ and corresponding target values $\{t_i\}$, $T=\bigcup_{i=1}^{n}U_i$, where $n$ is the number of target regions, then the $y^*$ is defined as:
\begin{equation}
\begin{aligned}
&\forall k\in[1,n], \forall (i,j) \in U_k,\,\, y^*_{i,j} = t_k \\
&\forall (i,j) \in O \setminus  T, \,\, y^*_{i,j} = \hat{y}_{i,j}.
\end{aligned}
\end{equation}
Since the adversary is accessible to parameters and gradients information of the target model under white-box settings, the attack can be formulated as the following optimization problem: 
\begin{equation}
\min_{\delta} \,\,  \mathcal{L}_{\text{total}}(x,\delta,y^*,f) \quad \text{s.t.} \,\, \|\delta\|_\infty \leq \tau,
\end{equation}
where $\mathcal{L}_{total}$ represents the total loss function to be optimized.

\subsection{Perturbations Initialization}
\label{sec:ptb-init}

Existing methods primarily employ two perturbation strategies: applying perturbations to the entire audio segment or applying them only to localized segments. The former tends to achieve higher attack effectiveness because it introduces more significant changes to the audio, but it often suffers from poor stealth, resulting in significant audio quality degradation and unintended alterations to the non-target region’s output, which exposes the attack behavior. The latter introduces fewer perturbations but lacks control over the entire audio context. For SED, which is highly sensitive to temporal information, this approach may fail to guarantee attack effectiveness unless improved.

\AttackName adopts the localized perturbation strategy and proposes improvements to it. The localized perturbations cover only the time span within the target region \(T\) as defined in Section~\ref{sec:formulation}. If multiple target regions \( T_i \) overlap in time, they share a common perturbation. Ultimately, these perturbations are denoted as \(\delta\), which introduces the minimum distortions while fully covering the target region. This approach takes into account the SED model's sensitivity to temporal information, suggesting that perturbations should ideally be applied only to the target region. However, since non-target region also influence the target region, further optimization steps will be introduced in Section~\ref{sec:optimize} to ensure that the attack does not disrupt non-target region.

\subsection{Optimization Process}\label{sec:optimize}

The optimization goal is to iteratively adjust the perturbations \( \delta \) added to the input audio in order to modify the target region \( T \) of the model output, while minimizing unintended modifications in the non-target region \( O \setminus T \). Intuitively, by using the commonly used BCE loss in SED tasks, this optimization problem can be formulated as:
\begin{equation}
\min_{\delta} \mathcal{L}_{\text{BCE}}(f(x + \delta), y^*), \quad \text{s.t.} \|\delta\|_\infty \leq \tau.
\end{equation}
This represents the traditional adversarial attack optimization objective, which can achieve certain effects. However, as analyzed in Section~\ref{sec:intro}, single-objective optimization alone struggles to simultaneously ensure effective attacks on the target region while keeping the non-target region unaffected. Empirically, we have observed that such attacks often unintentionally alter the non-target region, causing it to be misclassified as other events. 

To address this issue, we propose a dual-objective optimization approach that separately optimizes the target and non-target regions distinctly. A weighting factor is used to balance these two objectives, aiming to maintain the attack effectiveness on the target region while reducing unintended impacts on the non-target region. Specifically, we reformulate the optimization objective as:
\begin{equation}
\min_{\delta} \mathcal{L}_{\text{adv}}(f(x + \delta), y^*) + \alpha \cdot \mathcal{L}_{\text{pre}}(f(x + \delta), f(x)), \quad \text{s.t.} \|\delta\|_\infty \leq \tau,
\end{equation}
where \( \mathcal{L}_{\text{adv}} \) represents the \textit{adversarial loss}, which is used to optimize the target region of the output to the desired event, \( \mathcal{L}_{\text{pre}} \) represents the \textit{preservation loss}, which is used to maintain the non-target region unchanged, \(\alpha\) is a scaling factor that controls the balance between the adversarial loss and the preservation loss. Therefore, 
we define our total loss $\mathcal{L}_{total}$ as the weighted summation of the adversarial and preservation losses. The specific design of these two losses is as follows.

\noindent\textbf{Adversarial Loss.}
The adversarial loss \( \mathcal{L}_{\text{adv}} \) focuses on modifying the output in the target region \( T \) to match the desired target label matrix \( y^* \). Specifically, for each time-event pair \( (i,j) \in T \), the model output is updated to approach the target label \( y^*_{i,j} \). The adversarial loss is defined as:
\begin{equation}
\mathcal{L}_{\text{adv}}(f(x + \delta), y^*) = - \sum\nolimits_{(i,j) \in T} ( y^*_{i,j} \log(f(x + \delta)_{i,j}) + (1 - y^*_{i,j}) \log(1 - f(x + \delta)_{i,j}) ),
\end{equation}
where \( f(x + \delta)_{i,j} \) is the output of the model at time step \( i \) for event class \( j \) after applying the perturbation \( \delta \), and \( y^*_{i,j} \) is the target label for that pair. This loss term ensures that the model output in the target region aligns with the desired outcome.

\noindent\textbf{Preservation Loss.}
The preservation loss \( \mathcal{L}_{\text{pre}} \) is designed to minimize changes in the non-target region \( O \setminus T \), ensuring that the adversarial perturbation does not unintentionally modify the model output in these areas. For each time-event pair \( (i,j) \in O \setminus T \), we aim to preserve the original output \( f(x)_{i,j} \). The preservation loss is defined as:
\begin{equation}
\begin{aligned}
\mathcal{L}_{\text{pre}}(f(x &+ \delta), f(x)) = - \sum\nolimits_{(i,j) \in O \setminus T} ( f(x)_{i,j} \log(f(x + \delta)_{i,j}) \\
&+ (1 - f(x)_{i,j}) \log(1 - f(x + \delta)_{i,j}) ).
\end{aligned}
\end{equation}
This loss term ensures that the perturbation does not lead to significant deviations in the non-target region's classification, thereby improving the stealthiness of the attack.

The optimization of perturbations is updated using I-FGSM~\cite{kurakin2018adversarial}:
\begin{equation}
\delta \leftarrow \delta - \beta \cdot \text{sign}(\nabla_{\delta} \mathcal{L}_{total}(x, \delta, y^*, f)),
\end{equation}
where \(\beta\) is the learning rate for updating the perturbation.

After optimization, the perturbations are applied to the original audio to generate the adversarial example, which is then saved for evaluation. The perturbations are constrained within a defined range to avoid excessive distortion and maintain stealth.

\section{Experiments}\label{sec:experiments}
In this section, we present our experimental setup, analyze attack performance analyses, conduct ablation studies, and examine mel-spectrum and post-processed outputs analyses to evaluate the effectiveness and precision aspect of existing methods and our \AttackName. All adversarial examples were generated on an experimental platform consisting of 2 NVIDIA GeForce RTX 4090 GPUs running Ubuntu 18.04 (64-bit). The extensive experiments are provided in Appendix~\ref{app:extensive_expr}.

\subsection{Experiment Setup}

\textbf{Datasets.} We evaluate on two benchmark datasets for sound event detection: DESED~\citep{turpault2019sound} with 10-second audio clips and TUT-SED~\citep{mesaros2017tut} with 1--3 minute segments. For testing, we randomly select 1000 samples, segmenting TUT-SED into 25-second clips for efficiency.

\textbf{Target Models.} We select two models: a CRNN~\citep{adavanne2017sound}, a representative traditional architecture and DCASE-2017 winner, trained using the authors' code, and ATST-SED~\citep{shao2024fine}, a state-of-the-art Transformer-based model, using its fine-tuned version.

\textbf{Parameter Settings.} Target model parameters follow official settings. For attacks, we use Adam~\citep{kingma2014adam} with a learning rate of $1\times10^{-3}$ and 3-second event manipulations. For CRNN, we set $\alpha=10$, $\tau=0.02$; for ATST-SED, $\alpha=50$, $\tau=0.05$, tuned via ablation studies discussed in Section~\ref{sec:ablation}. Other parameters align with baselines.

\textbf{Baselines.} The C\&W attack~\cite{carlini2018audio} introduces perturbations globally across the entire audio segment while constraining the maximum perturbation. FAAG~\cite{miao2022faag} applies perturbations to localized audio segments, achieving effects comparable to global perturbations. ARO~\cite{jin2025boosting} leverages acoustic features to enhance the transferability of adversarial examples across different models.

\textbf{Evaluation Metrics.} We propose a composite metric system combining three novel metrics—Edit Precision (EP), Attack Success Rate (ASR), Unintended Editing Rate (UER)—with the existing Signal-to-Noise Ratio (SNR). EP measures the accuracy of adversarial modifications, balancing successful edits in target regions with preservation of non-target frames. ASR evaluates the effectiveness of altering intended frames. UER quantifies unintended modifications in non-target regions, with lower values indicating greater precision. SNR assesses perturbation imperceptibility, with higher values indicating less noticeable distortions. Detailed definitions are provided in Appendix~\ref{app:evaluation_metrics}.

\subsection{Attack Performance Analyses}\label{sec:main-exp}

\textbf{Single-Target Manipulation Evaluation.}Table~\ref{tab:single} compares attack performance under single-target manipulation scenarios. C\&W achieves the highest ASR, reaching 91.30\% on CRNN and 99.11\% on ATST-SED, but at the cost of poor EP, which drops to 87.84\% and 96.51\%, respectively. It also suffers from high UER, increasing to 12.23\% on CRNN and 3.57\% on ATST-SED, while severely degrading audio quality with an SNR of 12.96 dB and 9.32 dB. FAAG minimizes unintended manipulations, achieving UER values of 2.13\% on CRNN and 0.52\% on ATST-SED, and occasionally surpasses our method in EP on CRNN. However, its ASR remains low, at only 76.65\% on CRNN and 50.86\% on ATST-SED, making it unreliable for adversarial attacks. ARO achieves the highest ASR, reaching 92.79\% on CRNN and 99.20\% on ATST-SED, but suffers from the lowest EP at 85.59\% and 96.53\%, respectively, and the highest UER on CRNN at 14.56\%, alongside poor audio quality with SNR values of 10.00 dB and 7.28 dB.

\begin{table*}[h]
\small
\centering
\caption{Attack performance under single-target manipulation scenarios.}
\begin{tabular}{c|cccc|cccc}
\toprule
\multirow{2}{*}{Attack Method} & \multicolumn{4}{c|}{CRNN} & \multicolumn{4}{c}{ATST-SED} \\
\cmidrule(lr){2-5} \cmidrule(lr){6-9}
 & EP ($\uparrow$) & ASR ($\uparrow$) & UER ($\downarrow$) & SNR ($\uparrow$) & EP ($\uparrow$) & ASR ($\uparrow$) & UER ($\downarrow$) & SNR ($\uparrow$) \\
\midrule
C\&W & 87.84\% & 91.30\% & 12.23\% & 12.96dB & 96.51\% & 99.11\% & 3.57\% & 9.32dB \\
FAAG & \textbf{97.45\%} & 76.65\% & \textbf{2.13\%} & 19.42dB & 98.03\% & 50.86\% & \textbf{0.52\%} & \textbf{13.55dB} \\
ARO & 85.59\% & \textbf{92.79\%} & 14.56\% & 10.00dB & 96.53\% & \textbf{99.20\%} & 3.55\% & 7.28dB \\
\AttackName & 94.56\% & 81.81\% & 5.18\% & \textbf{20.16dB} & \textbf{99.11\%} & 91.77\% & 0.67\% & \textbf{13.55dB} \\
\bottomrule
\end{tabular}
\label{tab:single}
\end{table*}

In contrast, our method balances attack effectiveness and precision, attaining high EP values of 94.56\% on CRNN and 99.11\% on ATST-SED while maintaining competitive ASR at 81.81\% and 91.77\%, respectively. It also achieves lower UER, reducing unintended modifications to 5.18\% on CRNN and 0.67\% on ATST-SED, while better preserving audio quality with an SNR of 20.16 dB and 13.55 dB. These results confirm that our approach effectively manipulates target events while minimizing collateral impact.

Interestingly, despite being an older model, CRNN exhibits greater robustness against adversarial attacks than ATST-SED, making it harder to maintain a low UER. This could stem from its model aggregation strategy, which enhances resilience. Conversely, ATST-SED’s superior event recognition and feature disentanglement allow more precise input perturbations, making it more susceptible to adversarial modifications. Despite architectural and robustness differences, our method consistently improves attack performance on both, demonstrating adaptability. By balancing attack effectiveness and stealthiness across different SED architectures, our approach proves broadly applicable to adversarial attacks on SED systems.

\textbf{Multi-Target Manipulations Evaluation.} We evaluate performance under more complex manipulations conditions and investigate whether a high UER in single-target scenarios leads to a higher UER in multi-target scenarios. Table~\ref{tab:multi} presents the results of baselines and our method under three randomly applied edits. To minimize event overlap, we reduced the duration of each edited segment from 3 seconds to 2 seconds.

\begin{table*}[h]
\small
\centering
\caption{Attack performance under multi-target manipulations scenarios.}
\begin{tabular}{c|cccc|cccc}
\toprule
\multirow{2}{*}{Attack Method} & \multicolumn{4}{c|}{CRNN} & \multicolumn{4}{c}{ATST-SED} \\
\cmidrule(lr){2-5} \cmidrule(lr){6-9}
 & EP ($\uparrow$) & ASR ($\uparrow$) & UER ($\downarrow$) & SNR ($\uparrow$) & EP ($\uparrow$) & ASR ($\uparrow$) & UER ($\downarrow$) & SNR ($\uparrow$) \\
\midrule
C\&W & 87.66\% & \textbf{94.37\%} & 12.62\% & 10.83dB & 91.35\% & \textbf{99.85\%} & 9.22\% & 6.69dB \\
FAAG & \textbf{95.07\%} & 61.79\% & \textbf{3.56\%} & 16.26dB & 96.33\% & 51.20\% & \textbf{0.80\%} & \textbf{10.13dB} \\
ARO & 86.27\% & 90.28\% & 13.89\% & 9.83dB & 91.95\% & 99.67\% & 8.56\% & 6.71dB \\
\AttackName & 94.97\% & 82.96\% & 4.53\% & \textbf{16.77dB} & \textbf{98.84\%} & 96.57\% & 1.06\% & 9.84dB \\
\bottomrule
\end{tabular}
\label{tab:multi}
\end{table*}

The results show that under multi-target manipulations scenarios, C\&W’s UER increases slightly on CRNN, rising from 12.23\% to 12.62\%, whereas our method’s UER decreases slightly from 5.18\% to 4.53\%. On ATST-SED, C\&W’s UER rises significantly from 3.57\% to 9.22\%, while our method maintains a low UER, increasing only slightly from 0.67\% to 1.06\%. FAAG maintains the lowest UER at 3.56\% on CRNN and 0.80\% on ATST-SED, and achieves a high EP of 95.07\% on CRNN, but its ASR drops to 61.79\% and 51.20\%, rendering it ineffective for attacks. ARO’s UER increases significantly to 13.89\% on CRNN and 8.56\% on ATST-SED, with low EP values of 86.27\% and 91.95\%, despite a high ASR of 90.28\% and 99.67\%. Despite these differences, a general trend emerges: C\&W accumulates unintended manipulations more rapidly in multi-target settings, whereas our method exhibits greater stability. As a result, C\&W’s EP declines from 87.84\% to 87.66\% on CRNN and from 96.51\% to 91.35\% on ATST-SED, while our method maintains higher EP values of 94.97\% and 98.84\%, respectively. These findings highlight the importance of minimizing UER, as increased unintended manipulations not only degrade precision but also make attacks more detectable.

\subsection{Ablation Study} \label{sec:ablation}

To better understand the key factors influencing attack performance, we conduct an ablation study focusing on the contribution of preservation loss and the impact of hyperparameter selection. We specifically analyze the choice of these factors to achieve a more effective and precise attack.

\textbf{Contribution of Preservation Loss.} Preservation loss reduces unintended modifications in the non-target region while maintaining attack effectiveness. The parameter $\alpha$ regulates the balance between preservation and adversarial loss, influencing the trade-off between ASR and UER. To assess its impact, we evaluate eight $\alpha$ values, as shown in Figure~\ref{fig:alpha_ablation}. The results confirm that minimizing UER while maintaining a sufficiently high ASR leads to a more precise attack. For the CRNN model, setting $\alpha$ to 10 achieves an ASR above 80\% with minimal unintended manipulations, whereas for ATST-SED, $\alpha=50$ ensures an ASR above 90\% while keeping UER low. These findings demonstrate that preservation loss improves attack precision by constraining non-target perturbations. Additionally, $\alpha$ can be adjusted for scenarios requiring a stricter trade-off, such as prioritizing ASR over stealthiness.

\textbf{Hyperparameter Analysis.} Selecting an appropriate perturbation constraint is crucial for optimizing attack performance. The parameter \( \tau \) controls the \( \ell_\infty \) boundary of the perturbation \( \delta \), directly affecting attack effectiveness. To identify the optimal \( \tau \), we reference prior works and evaluate ten values ranging from 1 to 0.001. As shown in Figure~\ref{fig:tau_albation}, for the CRNN model, \( \tau=0.02 \) provides the best balance between effectiveness and stealthiness, while for ATST-SED, \( \tau=0.05 \) achieves a high ASR with moderate perturbation levels. These results validate the importance of carefully tuning \( \tau \) to achieve precise yet inconspicuous attacks.

\begin{minipage}{0.48\textwidth}
  \begin{figure}[H]
    \centering
    \includegraphics[width=\linewidth]{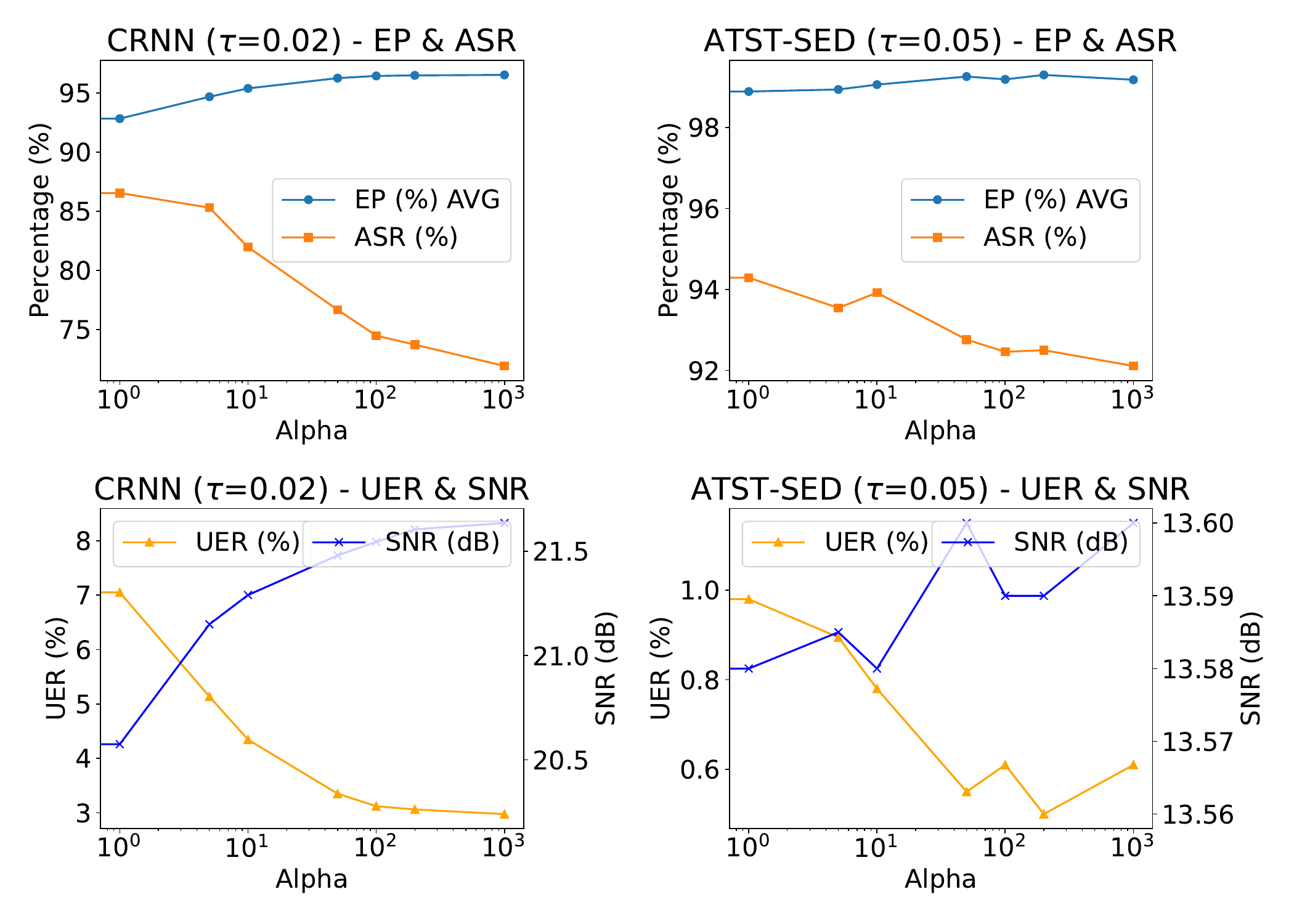}
    \caption{Effect of different $\alpha$ values on attack performance. There exists a trade-off between ASR and UER. We selected the $\alpha$ value that maximizes EP while ensuring sufficient ASR.}
    \label{fig:alpha_ablation}
  \end{figure}
\end{minipage}%
\hfill
\begin{minipage}{0.48\textwidth}
  \begin{figure}[H]
    \centering
    \includegraphics[width=\linewidth]{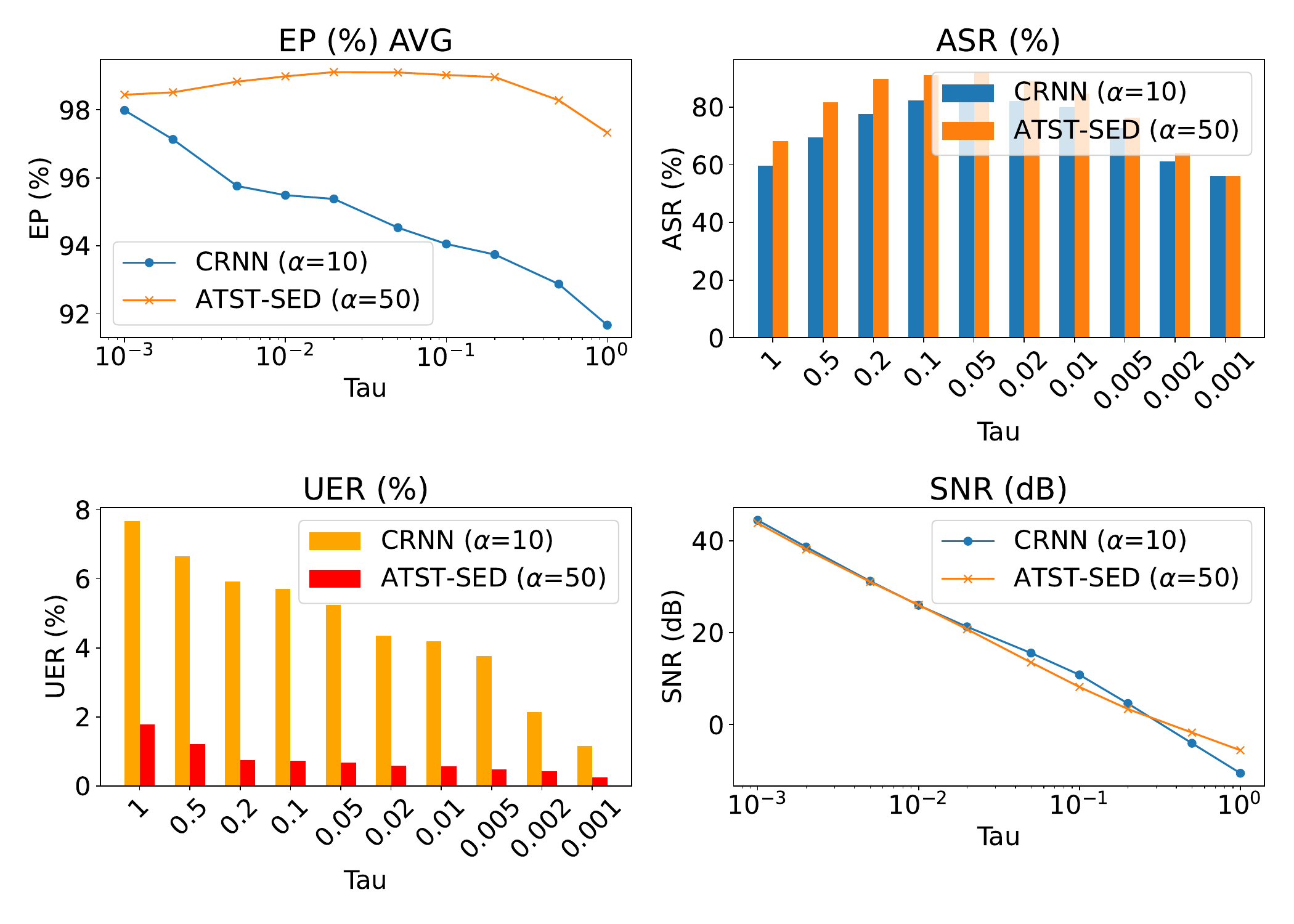}
    \caption{Effect of different $\tau$ values on attack performance. The trends of EP and ASR exhibit strong similarity, allowing $\tau$ selection to maximize EP with adequate ASR.}
    \label{fig:tau_albation}
  \end{figure}
\end{minipage}

\begin{figure*}[h]
    \centering
    \includegraphics[width=\linewidth]{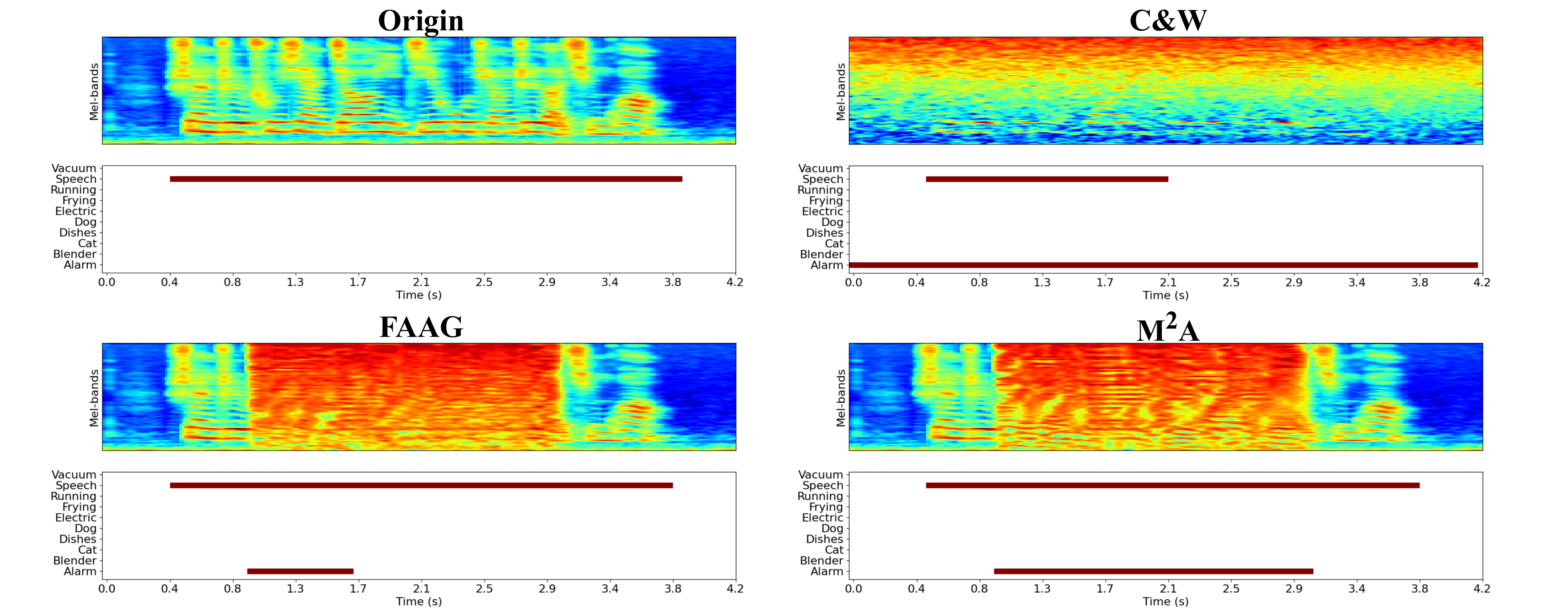}
    \caption{Mel-spectrum and post-processed output of adversarial examples generated by different methods. The objective is to insert `alarm' from 1.0 to 3.0s. While C\&W successfully achieves the attack objective, it introduces unintended modifications to the non-target region. FAAG, on the other hand, fails to fully achieve the intended modification in the target region. In contrast, our method precisely manipulates the target event while preserving the non-target region.
    }
    \label{fig:spec-output}
\end{figure*}

\subsection{Mel-Spectrum and Post-Processed Outputs Analyses}

To visually demonstrate the impact of different attack methods on audio signals and highlight variations in effectiveness and precision, we analyze spectral representations and post-processed detection outputs of adversarial examples. As shown in Figure~\ref{fig:spec-output}, C\&W causes the most severe quality loss, unintentionally altering `speech' and `alarm' events in the non-target region, making it imprecise. FAAG introduces minimal perceptible distortion but fails to fully activate all `alarm' instances from 1.0 to 3.0s, rendering it ineffective. In contrast, our method successfully achieves the attack objective without altering the non-target region, preserving audio quality and demonstrating both effectiveness and precision.

While C\&W effectively modifies the target region, it disrupts the non-target region. FAAG minimizes unintended modifications but lacks sufficient attack effectiveness, leading to incomplete target event manipulation. Our method balances attack effectiveness and minimal collateral effects. Although no method achieves perfect editing precision for all samples, our approach consistently delivers the best performance, reinforcing conclusions from Section~\ref{sec:main-exp}.

\subsection{Countermeasures Evaluation}
\label{sec:countermeasure}

To evaluate the robustness of our proposed \AttackName attack against existing defense mechanisms, we adapted several strategies, widely validated in speech-related tasks, to mitigate adversarial perturbations in sound event detection (SED). This analysis aims to assess the effectiveness of these defenses in countering \AttackName and to determine the attack's resilience under such countermeasures. Specifically, we implemented three defense strategies: Downsampling~\cite{li2024attack}, Gaussian Noise~\cite{liu2022friendly}, and Local Smoothing~\cite{naseer2019local} (including mean and median smoothing). Detailed setups for these defenses are provided in Appendix~\ref{app:defense_setup}.

We evaluated these defenses against \AttackName using four metrics: Edit Precision (EP), Attack Success Rate (ASR), Unintended Edit Rate (UER), and Signal-to-Noise Ratio (SNR). Table~\ref{tab:defense} reports the results, which indicate that down sampling and local smoothing effectively reduce ASR, particularly for the ATST-SED model, though they incur trade-offs in SNR and audio quality. Gaussian noise, however, shows limited effectiveness, often increasing UER and reducing attack precision. Notably, \AttackName retains significant efficacy against these defenses, underscoring its robustness.

\begin{table*}[h]
\centering
\small
\addtolength{\tabcolsep}{-2pt}
\caption{Performance of defense mechanisms against \AttackName.}
\begin{tabular}{c|cccc|cccc}
\toprule
\multirow{2}{*}{Defense Method} & \multicolumn{4}{c|}{CRNN} & \multicolumn{4}{c}{ATST-SED} \\
\cmidrule(lr){2-5} \cmidrule(lr){6-9}
 & EP ($\uparrow$) & ASR ($\downarrow$) & UER ($\downarrow$) & SNR ($\uparrow$) & EP ($\uparrow$) & ASR ($\downarrow$) & UER ($\downarrow$) & SNR ($\uparrow$) \\
\midrule
N/A & 94.56\% & 81.81\% & 5.18\% & 20.16dB & 99.11\% & 91.77\% & 0.67\% & 13.55dB \\
Down Sampling & 94.60\% & 72.01\% & 4.94\% & 20.42dB & 97.53\% & 50.85\% & 1.04\% & 7.04dB \\
Gaussian Noise & 90.53\% & 80.26\% & 9.26\% & 15.01dB & 98.31\% & 80.87\% & 1.15\% & 12.08dB \\
Mean Smoothing & 94.64\% & 70.57\% & 4.87\% & 22.29dB & 98.05\% & 62.85\% & 0.87\% & 10.08dB \\
Median Smoothing & 93.98\% & 76.48\% & 5.66\% & 19.19dB & 97.65\% & 56.76\% & 1.10\% & 9.79dB \\
\bottomrule
\end{tabular}
\label{tab:defense}
\end{table*}

\section{Discussion and Limitations}
\label{sec:discuss}

While this study exposes critical weaknesses in SED systems, it also underscores the necessity of developing more robust defense mechanisms. One promising direction is leveraging ensemble-based architectures, as our results indicate that the evaluated CRNN exhibits greater resistance to adversarial attacks, likely due to its model aggregation strategy. Additionally, enhancing context dependencies in SED models could help mitigate adversarial manipulations while preserving detection accuracy. Addressing these vulnerabilities is crucial for ensuring the reliability of SED in safety-critical applications, promoting more secure and trustworthy audio-based surveillance systems.

\section{Conclusion}
In this work, we present the first targeted adversarial attack framework for polyphonic SED, \AttackName, achieving high attack precision through localized perturbations and a novel preservation loss constraint while ensuring sufficient attack effectiveness. Our method enables precise event manipulations while effectively minimizing unintended modifications in the non-target region, as demonstrated on TUT-SED and DESED using ATST-SED and CRNN models. Our findings highlight that SED exhibits stronger context dependencies than other audio tasks, making adversarial attacks more challenging. Despite this, our attack effectively exploits inherent vulnerabilities in SED models, revealing potential security risks in safety-critical applications.

\bibliographystyle{plain}
\bibliography{ref}

\clearpage
\appendix

\section{\AttackName Algorithm}

The pseudocode of the \AttackName\ algorithm is shown in Algorithm~\ref{alg:MM}. Perturbations are first initialized as described in Section~\ref{sec:ptb-init}. The algorithm proceeds iteratively, computing the adversarial loss $\mathcal{L}_{adv}$ and preservation loss $\mathcal{L}_{pre}$ for the target and non-target regions, respectively. During optimization, the preservation loss is weighted by a factor $\alpha$, and the total loss $\mathcal{L}_{total} = \mathcal{L}_{adv} + \alpha \mathcal{L}_{pre}$ is used to compute gradients for optimization. After iteration, perturbations are added to the original audio to produce the target adversarial example.

\begin{algorithm}[H]
\caption{\AttackName}
\label{alg:MM}
\begin{algorithmic}[1]
\State \textbf{Input:} Target SED model $f$, original audio $x$, target label matrix $y^*$, target region $T$, scaling factor $\alpha$, learning rate $\beta$, maximum iteration number $N$, perturbations' threshold $\tau$.
\State \textbf{Output:} Optimized audio adversarial example $x'$.
\State Initialize perturbations $\delta$ based on $T$ and $\tau$;
\For{$i \leftarrow 1$ \textbf{to} $N$}
    \State $x' \gets x + \delta$;
    \State Calculate adv loss $\mathcal{L}_{adv}$;
    \State Calculate preservation loss $\mathcal{L}_{pre}$;
    \State $\mathcal{L}_{total}\gets\mathcal{L}_{adv}+\alpha\cdot\mathcal{L}_{pre}$;
    \State $\delta \gets \delta - \beta \cdot \text{sign}(\nabla_{\delta}\mathcal{L}_{total}(x,\delta,y^*,f))$;
\EndFor
\State $x' \gets x + \delta$;
\State \textbf{return} $x'$.
\end{algorithmic}
\end{algorithm}

\section{Experimental Details}
\label{app:exp_details}

\subsection{Evaluation Metrics}\label{app:evaluation_metrics}

We evaluate both attack effectiveness and precision using the following metrics:

\textbf{Editing Precision (EP)} quantifies how accurately the adversarial output aligns with the intended target while minimizing unintended modifications. Inspired by the F1-score \citep{fisher1936use, bilen2020framework}, EP accounts for both successful modifications in the target region and preservation of non-target frames.
\begin{equation}
\text{EP} = \frac{SE + NE}{SE + FE + UE + NE},
\end{equation}
where Successful Editing (SE) represents correctly modified frames in the target region, Failed Editing (FE) denotes frames that remain unchanged despite the attack, Unintended Editing (UE) counts unintended modifications in the non-target region, and Not Editing (NE) refers to non-target frames that remain unaltered. A higher EP indicates a more precise and stealthy attack.

\textbf{Attack Success Rate (ASR)} measures the effectiveness of modifying target frames:

\begin{equation}
\text{ASR} = \frac{SE}{SE + FE}.
\end{equation}
A higher ASR suggests greater success in altering the intended frames.

\textbf{Unintended Editing Rate (UER)} quantifies the degree of undesired modifications:
\begin{equation}
\text{UER} = \frac{UE}{UE + NE}.
\end{equation}
A lower UER indicates fewer unintended manipulations in the non-target region.

\textbf{Signal-to-Noise Ratio (SNR)} evaluates perturbation imperceptibility, with higher values indicating less noticeable distortions. It is defined as:
\begin{equation}
\text{SNR} = 10 \log_{10} \left( \frac{\sum_{i=1}^{N} x_i^2}{\sum_{i=1}^{N} (x_i - x'_i)^2} \right),
\end{equation}
where $x$ represents the original audio signal of $N$ samples, and $x'$ denotes the adversarially perturbed audio. A higher SNR suggests a less perceptible attack.

These metrics comprehensively assess attack performance, ensuring precise event manipulation with minimal unintended effects.

\subsection{Countermeasures Setup}
\label{app:defense_setup}

To thoroughly evaluate the robustness of our proposed \AttackName attack, we implemented three defense mechanisms adapted from strategies widely validated in speech-related tasks. These defenses aim to mitigate adversarial perturbations in sound event detection (SED) by disrupting or masking the adversarial modifications introduced by \AttackName. Below, we detail each defense method, referencing their official implementations and describing our specific configurations.

\begin{itemize}[leftmargin=*]
    \item \textbf{Down Sampling~\cite{li2024attack}}: We downsample audio from its original rate (e.g., 44.1 kHz) to a lower rate (e.g., 16 kHz) to remove high-frequency adversarial perturbations, then upsample via linear interpolation.

    \item \textbf{Gaussian Noise~\cite{liu2022friendly}}: Random Gaussian noise with a standard deviation of 0.01 is added to the audio to mask adversarial perturbations. This straightforward defense mechanism can successfully mitigate specific fragile adversarial perturbations, notably those based on suffixes or prefixes, in an effective manner.

    \item \textbf{Local Smoothing~\cite{naseer2019local}}: We apply mean and median smoothing using a sliding window of size equal to 3. Mean smoothing averages samples within the window via one-dimensional convolution, while median smoothing selects the median value, preserving edge features.
\end{itemize}

\section{Extensive Experiments}
\label{app:extensive_expr}

\begin{figure}[h]
    \centering
    \includegraphics[width=0.65\linewidth]{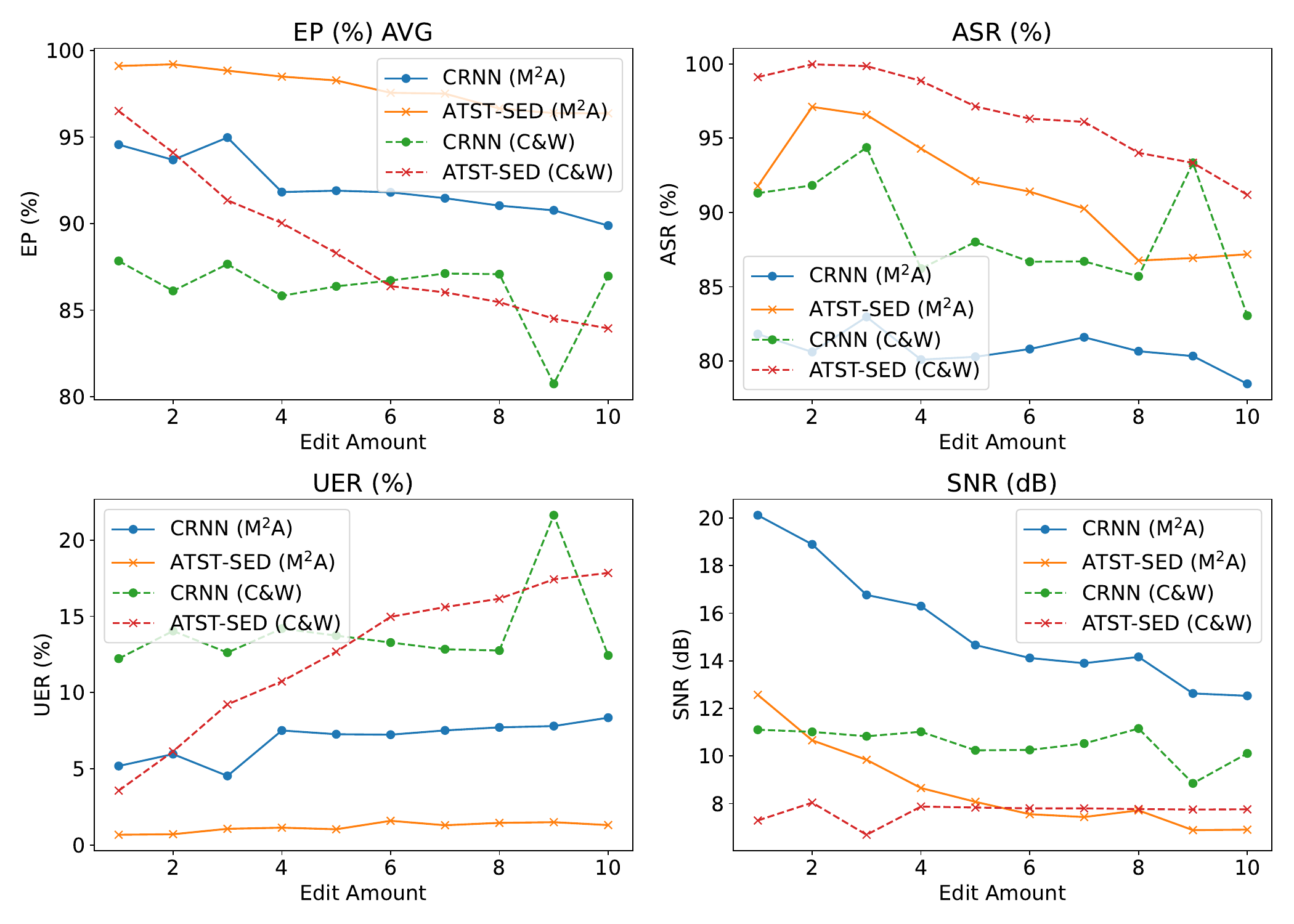}
    \caption{The trend of attack performance changes when the number of edits increases from 1 to 10.}
    \label{fig:edit-amount}
\end{figure}

To further validate our method under arbitrary manipulations conditions, we conduct additional experiments with manipulation counts ranging from 1 to 10. As shown in Figure~\ref{fig:edit-amount}, EP decreases moderately while UER increases gradually. However, even after 10 edits, EP remains above 89.89\% on CRNN and 96.38\% on ATST-SED, confirming that our method maintains high precision across multiple edits, making it well-suited for real-world attack scenarios.

\end{document}